\documentclass[12pt]{article}
\usepackage{amsmath,amssymb,amsthm,amsxtra,overpic,bbm,bm,epsfig,ulem,multirow}
\usepackage{color,cite}
\usepackage{epstopdf}
\usepackage{booktabs}
\usepackage{footnote}
\usepackage{rotating}
\usepackage{hyperref}
\usepackage{url}
\usepackage{cite}
\usepackage{tikz}
\usetikzlibrary{arrows,shapes,snakes,automata,backgrounds,petri}
\usetikzlibrary{decorations.pathmorphing}
\usetikzlibrary{decorations.markings}
\usepackage{lineno}
\usepackage{graphicx}
\usepackage{subcaption}
\usepackage{diagbox}

\def\thefootnote{\fnsymbol{footnote}}

\usepackage{array}

\vspace{0.2cm}

\begin{document}

\begin{center}
{\Large\bf Neutral-current background induced by atmospheric neutrinos at large liquid-scintillator detectors: III. Comprehensive prediction for low energy neutrinos} 
\end{center}


\vspace{0.2cm}

\begin{center}
{\bf Jie Cheng$^a$~\footnote{Co-first author, these authors contributed equally to this work.}, Min Li$^b$~\footnote{Co-first author, these authors contributed equally to this work.}, Yu-Feng Li$^{b,c}$~\footnote{Email: liyufeng@ihep.ac.cn}, Gao-Song Li$^b$, Hao-Qi Lu$^b$~\footnote{Email: luhq@ihep.ac.cn}, Liang-Jian Wen$^b$ }
\\
\vspace{0.2cm}
{$^a$School of Nuclear Science and Engineering, North China Electric Power University, Beijing 102206, China}\\
{$^b$Institute of High Energy Physics, Chinese Academy of Sciences, Beijing 100049, China}\\
{$^c$School of Physical Sciences, University of Chinese Academy of Sciences, Beijing 100049, China}
\end{center}

\vspace{1.5cm}

\begin{abstract}

Atmospheric neutrinos play a vital role in generating irreducible backgrounds in liquid-scintillator (LS) detectors via their neutral-current (NC) interactions with $^{12}$C nuclei. These interactions may affect a wide range of research areas from the MeV to GeV energy range, such as the reactor and geo neutrinos, diffuse supernova neutrino background (DSNB), dark matter, and nucleon decay searches. In this work, we extend our preceding paper~\cite{Cheng:2020aaw}, by conducting a first-time systematic exploration of NC backgrounds as low as the MeV region of reactor and geo neutrinos. We utilize up-to-date neutrino generator models from GENIE and NuWro, a TALYS-based nuclear deexcitation package and a GEANT4-based detector simulation toolkit for our complete calculation. Our primary focus is to predict the NC background for experimental searches of inverse-beta-decay signals below the 100 MeV visible energy. In order to have deeper understanding of the characteristics of atmospheric neutrino NC interactions in LS, we investigate the model dependence of NC background predictions by using various data-driven models, including the initial neutrino-nucleon interactions, nuclear ground-state structure, final-state interactions, nuclear deexcitation processes, and secondary interactions of final-state particles. 

\end{abstract}

\def\thefootnote{\arabic{footnote}}
\setcounter{footnote}{0}

\newpage


\section{Introduction}
\label{sec:intro}

The neutral-current (NC) interactions between atmospheric neutrinos ($\nu_{\text{atm}}$) and nuclei are important signatures of the large water-Cherenkov (wCh) and liquid-scintillator (LS) detectors, especially for the neutrinos in the energy range of (0.1 $-$ 20) GeV~\cite{Formaggio:2012cpf}. The products of $\nu_{\text{atm}}$ NC interactions include the daughter nuclei, nucleons, mesons, $\gamma$-rays and invisible neutrinos.
These final-state particles contribute to the signals of single, time-correlated double or triple-coincident events in either wCh or LS detectors. It offers an alternative way to explore the GeV neutrino interaction mechanism and enhance our understanding of these interactions~\cite{KamLAND:2022ptk,Super-Kamiokande:2019hga}. Meanwhile, the $\nu_{\text{atm}}$ NC interactions significantly contribute to the irreducible backgrounds, encompassing the searches for rare events such as the diffuse supernova neutrino background (DSNB)~\cite{Super-Kamiokande:2022cvw, Super-Kamiokande:2021jaq,KamLAND:2021gvi,Borexino:2019wln}, proton decay~\cite{Super-Kamiokande:2022egr, Super-Kamiokande:2020wjk,KamLAND:2015pvi,JUNO:2022qgr}, and neutrinos from the dark matter annihilation~\cite{KamLAND:2021gvi,Super-Kamiokande:2020sgt,Super-Kamiokande:2022ncz}. In addition, the observable energy range of the $\nu_{\text{atm}}$ NC interactions can overlap with the reactor neutrino energy observation window~\cite{DayaBay:2018yms,DoubleChooz:2019qbj,RENO:2018dro}, mimicking the reactor neutrino signal and further escalating irreducible backgrounds.
Future large liquid-scintillator (LS) detectors, exemplified by the Jiangmen Underground Neutrino Observatory (JUNO)~\cite{JUNO:2021vlw}, exhibit significant potential to attain sub-percent precision in measuring the neutrino oscillation parameters and determining neutrino mass ordering. This potential is from their capability to detect reactor $\bar{\nu}_e$ inverse-beta-decay (IBD) reactions, $\bar{\nu}_e\, + \, p \rightarrow e^{+} \, + \, n$, characterized by a prompt positron signal and a delayed neutron capture signal. LS detectors, featuring excellent neutron tagging, inherently offer high efficiency for the IBD event selection. Nonetheless, to achieve highly precise measurements of reactor $\bar{\nu}_e$, careful consideration of $\nu_{\text{atm}}$-$^{12}$C NC interactions is essential.

In our preceding paper~\cite{Cheng:2020aaw}, denoted as ``the preceding" paper, we have systematically calculated the NC interactions of $\nu_{\text{atm}}$-$^{12}$C in large LS detectors. For our calculation, we have utilized the $\nu_{\text{atm}}$ fluxes at the JUNO site from the Honda group~\cite{hondaflux, Honda:2015fha}, and determined rates and spectra for both inclusive and exclusive channels of $\nu_{\text{atm}}$-$^{12}$C NC interactions within the LS. In the the preceding paper, we have adopted a two-step calculation approach. Firstly, we have employed advanced neutrino interaction generators GENIE~\cite{Andreopoulos:2009rq} and NuWro~\cite{Golan:2012rfa} to calculate neutrino-$^{12}$C interactions. Subsequently, the TALYS~\cite{Koning:2005ezu} package is used to predict the deexcitation of the residual nucleus. This prediction method for the $\nu_{\text{atm}}$ NC background is extensively used in various physical topics within the JUNO collaboration, including reactor neutrinos to measure neutrino oscillation parameters~\cite{JUNO:2022mxj} and identify neutrino mass ordering~\cite{JUNO:2024jaw}, the prospects for detecting the DSNB~\cite{JUNO:2022lpc}, the annihilation of MeV dark matter in the galactic halo~\cite{JUNO:2023vyz} and invisible decay modes of neutrons~\cite{JUNO:2024pur}. 

Expanding on this basis, in the current study, we firstly integrate additional sophisticated neutrino generator models from the updated versions of GENIE (3.0.6) and NuWro (19.02). Moreover, by incorporating a GEANT4-based detector simulation~\cite{GEANT4:2002zbu} for the second interactions of particles propagating in the LS medium, we develop a three-fold strategy for predicting $\nu_{\rm atm}$ NC interactions. All models are employed to examine the characteristics of $\nu_{\text{atm}}$-$^{12}$C interactions, encompassing both the initial $\nu_{\text{atm}}$-nucleon interaction within $^{12}$C and final-state distributions to demonstrate model variations.
This investigation not only elucidates the influence of various models of initial nuclear states, final-state interactions (FSI), and the nuclear deexcitation process on the calculation of $\nu_{\text{atm}}$-$^{12}$C interactions, but also incorporates the effect of secondary interactions of final-state neutrons on the final tagged neutron multiplicity. Finally,
the preceding paper has delved into the implication of the $\nu_{\text{atm}}$ NC background concerning the searches for the DSNB and proton decay. In contrast, this work extends to much lower energies to
fit with the reactor and geo neutrino studies.
Furthermore, with the introduction of new data-driven models, a discussion on the NC background for DSNB is included, providing a comparative analysis with the predictions from previous models.

The structure of the remaining sections in this work is outlined as follows. In Sec.~\ref{sec:Calcalation}, we present a comprehensive overview of the calculation methodology, discussing the evolution of neutrino generator models from early to recent times, the TALYS-based deexcitation model, and the GEANT4-based detector simulation. Sec.~\ref{sec:NCchar} focuses on highlighting key characteristics derived from the simulation of the $\nu_{\text{atm}}$-$^{12}$C interaction, including discussions on the initial $\nu_{\text{atm}}$-nucleon interaction and final-state information. Moving to Sec.~\ref{sec:res}, we present the final results of the NC background. Finally, in Sec.~\ref{sec:sum}, we summarize our main results and conclude the paper.

\section{Methodology for $\nu_{\rm atm}$-$^{12}$C NC Interaction Prediction} \label{sec:Calcalation}

\begin{figure}[!tb]
    \centering
	\includegraphics[width=0.7\linewidth]{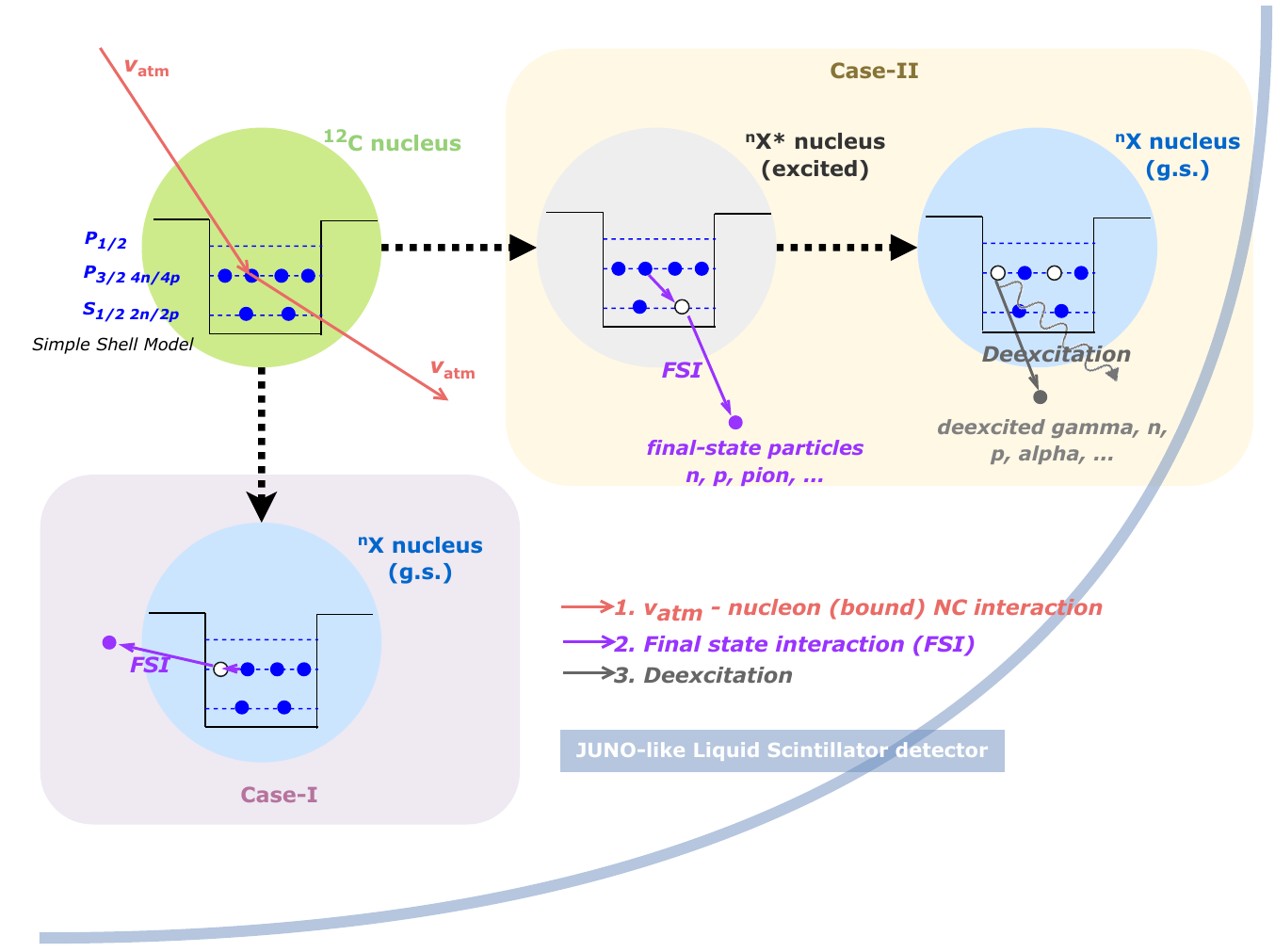}
    \caption{The diagram of the neutral current interactions between $\nu_{\rm atm}$ and $^{12}$C. The potential interaction processes following the initial $\nu_{\rm atm}$-nucleon interactions (depicted in pink) within the JUNO-like liquid scintillator detector, including final-state interactions (in purple) and deexcitation of the residual nucleus (in gray) are shown. A simple shell model accounts for the $^{12}$C nuclear structure.  The ``X" refers to the name of the residual nucleus and the ``g.s" denotes the ground state of the residual nucleus. }
	\label{fig:ncprocess}      
\end{figure}

Our current comprehension of the $\nu_{\rm atm}$ interaction mechanism mainly relies on model predictions, especially for NC interactions between the $\nu_{\rm atm}$ and nuclei. 
Considering the primary energy range of $\nu_{\rm atm}$ is around the GeV scale, the NC interaction mechanism between the $\nu_{\rm atm}$ and nuclei is exceptionally complex, presenting numerous models each with its own set of challenges. 
Figure~\ref{fig:ncprocess} illustrates the schematic procedure for calculating the NC interactions of $\nu_{\rm atm}$ with $^{12}$C. This includes the initial-state $\nu_{\rm atm}$-nucleon NC interaction and various aspects of nuclear effects.


For the initial-state interaction, as the energy of the incident neutrino increases from 100 MeV to 10 GeV or even higher, the dominant contribution to the cross-section typically originates from different interaction processes, including quasi-elastic scattering (QE), coherent and diffractive production (COH), nuclear resonance production (RES) and deep inelastic scattering (DIS). These processes are explained by using parameterization of form factors, with the primary uncertainty lying in the axial form factors.
For instance, in the current parameterized models of the NCQE interaction mechanism between the $\nu_{\rm atm}$ and nucleon, the axial mass $M_{\rm A}$ 
constitutes one of the most challenging parameters for experimental determination. Despite numerous experiments measuring $M_{\rm A}$, it exhibits substantial measurement uncertainties. Present measurements estimate the value of $M_{\rm A}$ in the range of around (1 $-$ 1.4) GeV~\cite{Kitagaki:1990vs,Bernard:2001rs,MiniBooNE:2010bsu}. 

Moreover, in the $\nu_{\rm atm}$-nucleus interactions, one should consider nuclear effects including the binding energy, nuclear ground state structure, many-body effects, FSI, deexcitation of the residual nucleus, and others. However, the nuclear effect models pose the following challenges:
\begin{itemize}
    \item The description of nuclear binding effects fluctuates across different models and requires additional exploration. The contemporary models, including Relativistic Fermi Gas (RFG), Local Fermi Gas (LFG), and Spectral Function (SF), are employed to characterize the nucleus ground state. However, these models may exhibit inconsistencies in describing the momentum distribution of nucleons bound within the nucleus.
    \item The inclusion of two-body current effects in QE, referred to as two-particle two-hole interactions ($2p2h$) or meson exchange currents (MEC), allows for the reproduction of MiniBooNE results with a smaller axial mass value, consistent with the world average obtained from other measurements~\cite{Nieves:2011pp, Martini:2009uj, Bodek:2011ps}. However, a direct measurement of the $2p2h$ interaction remains unattained, resulting in a considerable model-dependent uncertainty.
    \item  Nucleons involved in the neutrino initial-state interactions possess a high probability of interacting with surrounding nucleons within the nucleus, leading to FSI. However, discrepancies exist among various models in depicting this process. Moverover, post-FSI, the state of the residual nucleus and potential deexcitation processes remain incompletely addressed.
\end{itemize}

In addition to the primary $\nu_{\rm atm}$ interactions, predictions should also incorporate the secondary interactions, which occur when final-state particles interact within the detector after a primary interaction. These secondary interactions will influence the ultimate detector response and the particle identification in the detectors.

\subsection{Neutrino Generator Models}

In order to account for various contributions to the initial-state $\nu_{\rm atm}$-nucleon interactions and manage the nuclear effects in neutrino-nucleus interactions, we typically use sophisticated Monte Carlo neutrino event generators. Besides GENIE and NuWro, which have been used in our calculations to demonstrate neutrino generator model dependence, other generators like GiBUU~\cite{Buss:2011mx} and NEUT~\cite{Hayato:2009zz} are also available. These generators are carefully constructed and finely tuned according to available experimental data on interaction cross-sections. For example, predictions from GENIE and NuWro are evaluated against numerous experimental data like ANL, BNL, K2K, MiniBooNE and SciBooNE~\cite{Andreopoulos:2009rq,Juszczak:2005zs,Golan:2012wx}. Nevertheless, as previously mentioned, the lack of sufficient experimental data constraints on both initial-state neutrino-nucleon interactions and nuclear effects results in predictions being subject to a considerable level of uncertainty. Consequently, neutrino event generators provide numerous possibilities for the physics parameters and nuclear effect models, offering a wide array of potential neutrino generator models for simulating neutrino interactions with the nuclei. 

Given the existing focus on NC backgrounds within the visible energy range under 100 MeV, we adopt neutrino generator models that are predominantly centered on variations associated with the QE interactions between $\nu_{\rm atm}$ and $^{12}$C in LS. Specifically, eleven representative neutrino generator models are employed to evaluate the impact of different neutrino event generators, the axial mass $M_{\rm A}$ in QE, nuclear ground state, $2p2h$ effects, and FSI models on the predictions. Among these, six typical neutrino generator models are derived from GENIE (2.12.0) and NuWro (17.10). These models have been used in our preceding paper, with one from GENIE referred to as Model-G1 (G), and five from NuWro labeled as Model-N$i$ for $i = 1,2,3,4,5$.

GENIE and NuWro neutrino event generators persistently evolve their modern and universal event generator framework and tools to support neutrino experiments and provide novel comprehensive physics models for simulating neutrino interactions. For instance, GENIE has achieved a notable milestone with the release of GENIE v3~\cite{GENIE:2021npt}, which offers new physical tunes of neutrino interaction cross sections based on data from neutrino scattering experiments such as neutrino and antineutrino charge-current (CC) data on Carbon from MiniBooNE, T2K and MINER$\nu$A~\cite{GENIE:2022qrc,GENIE:2021zuu}. This signifies a shift from the prior approach in GENIE v2, which depends on a single "Default" model with optional modeling elements. NuWro (19.02) significantly improves the nucleon cascade model~\cite{Niewczas:2019fro}, which is able to reproduce nuclear transparency data of electrons, particularly in the energy range crucial for neutrino-nucleus scattering physics. 
Hence, our calculation incorporates the updated GENIE (3.0.6) and NuWro (19.02) for predicting the $\nu_{\rm atm}$ interactions. We employ five new neutrino generator models from GENIE (3.0.6) and NuWro (19.02), including three from GENIE designated as Model-G$i$ for $i = 2,3,4$, and two from NuWro labeled as Model-N6 and Model-N7.   
\renewcommand{\arraystretch}{1.2}
\begin{table}[!tb]
\centering
\caption{
Summary of the main features of models used in early and recent stages from Monte Carlo generators of neutrino interactions, GENIE~\cite{Andreopoulos:2009rq} and NuWro~\cite{Golan:2012rfa}, highlighting differences in generator versions and axial mass $M_{\rm A}$ for QE, usage of relativistic Fermi gas model with ``Bodek-Ritchie" modifications (BRRFG), local Fermi gas (LFG) model, or spectral function (SF) approach to nuclear structure, consideration of $2p2h$ interaction contribution to NCQE, and choice of FSI model.}
\begin{tabular}{llllll}
\hline \hline
Models &Generator   & $M_{\rm A}$   & Nuclear     & Inclusion   & FSI    \\
& (version) & [GeV] &model & of $2p2h$ & model \\
\hline
\multicolumn{6}{c}{\underline{Models used in preceding papers~\cite{Cheng:2020aaw,Cheng:2020oko}}} \\
Model-G1 (G) &GENIE (2.12.0) & 0.99 & BRRFG  & $\times$  & $h$A\\
Model-N1 &NuWro (17.10)  & 1.03 & LFG  & $\times$ &  Ref.~\cite{Golan:2012wx} \\
Model-N2 &NuWro (17.10)  & 0.99 & LFG  & $\times$ &  Ref.~\cite{Golan:2012wx} \\
Model-N3 &NuWro (17.10)  & 1.35 & LFG  & $\times$ &  Ref.~\cite{Golan:2012wx} \\
Model-N4 &NuWro (17.10)  & 0.99 & LFG  & \checkmark (TEM) & Ref.\cite{Golan:2012wx} \\
Model-N5 &NuWro (17.10)  & 0.99 & SF  & $\times$ &  Ref.~\cite{Golan:2012wx} \\
\multicolumn{6}{c}{\underline{New models added in this work}} \\
Model-G2 &GENIE (3.0.6)  & 0.96 & LFG  & \checkmark (EP) & $h$N \\
Model-G3 &GENIE (3.0.6)  & 0.96 & LFG  & \checkmark (EP) & $h$A \\
Model-G4 &GENIE (3.0.6)  & 0.96 & BRRFG  & \checkmark (EP) & $h$N \\
Model-N6 &NuWro (19.02)  & 1.03 & LFG  & $\times$ & Ref.~\cite{Niewczas:2019fro}\\
Model-N7 &NuWro (19.02)  & 1.03 & SF  & $\times$ & Ref.~\cite{Niewczas:2019fro}\\
\hline \hline
\end{tabular}
\label{table:models}
\end{table}

In Tab.~\ref{table:models}, we have compared the key features between these early adopted and recent models in our new calculation to illustrate the variations in neutrino interactions.
The principal characteristics of these neutrino generator models and the reasoning behind their selection are summarized as follows.
\begin{description}
    \item[Initial neutrino-nucleon interaction:] In the initial stage, we scrutinize the impact of the $M_{\rm A}$ value on the cross sections of QE interactions. Different models utilize unique choices for $M_{\rm A}$. For GENIE, Model-G1 employs a default setting of $M_{\rm A} = 0.99$ GeV, determined from the deuterium measurements~\cite{Kitagaki:1990vs}. Similarly, Model-G$i$ with $i = 2,3,4$ adopts a very close value of $M_{\rm A} = 0.96$ GeV and these new models incorporate the \texttt{G18\_10a/b} configurations and have been tuned by neutrino scattering data~\cite{GENIE:2021zuu}. In NuWro, three different $M_{\rm A}$ values are used: $M_{\rm A} = 0.99$ GeV for Model-N$i$ with $i = 1,4,5$, $M_{\rm A} = $ 1.03 GeV from the world average~\cite{Bernard:2001rs} for Model-N$i$ with $i = 2,6,7$, and $M_{\rm A} = $ 1.35 GeV from the MiniBooNE data~\cite{MiniBooNE:2010bsu} for Model-N3.
    \item[Nuclear model:] In GENIE and NuWro, various nuclear models such as RFG, LFG, and SF are available to determine the momentum and binding energy of the target nucleon. To investigate the impact of the uncertainty from nuclear models on the prediction, Model-G1/4 integrates the RFG model with "Bodek-Ritchie" modifications (BRRFG)~\cite{Bodek:1981wr}, while Model-G2/3 uses LFG. Regarding the NuWro generator, LFG and SF are used in Model-N$i$ for $i = 1,2,3,4,6$ and Model-N5/7, respectively.{\footnote{It is important to note that there is a typo in Refs.~\cite{Cheng:2020aaw,Cheng:2020oko}: Model-N$i$ for $i = 1,2,3,4$ employs the LFG instead of RFG.}}
    \item[Inclusion of the $2p2h$ effect:] Not all the models from GENIE and NuWro integrate $2p2h$ models to contemplate the two-body current effects on NCQE. For our selection, Model-N4 features the transverse enhancement (TEM) model of the meson exchange current to predict the $2p2h$ interactions, derived from fitting the electron scattering data~\cite{Bodek:2011ps}. In other NuWro models, a different $2p2h$ implementation is used only for CC interactions, by using the Nieves $2p2h$ model~\cite{Nieves:2011pp}, a theory-based model commonly employed in the neutrino analyses. GENIE (3.0.6) models adopt the empirical (EP) multi-nucleon model with implementations of the corresponding Valencia $2p2h$ models by Nieves et al.~\cite{Katori:2013eoa,Nieves:2011pp}.
    \item[FSI model:] Describing the effect of FSI is crucial yet challenging for neutrino-nucleus interactions. In the early models of GENIE (2.12.0), FSI is simulated using the intranuclear cascade (INC) $h$A model, an empirically tunable model primarily driven by hadron-nucleus scattering data~\cite{Andreopoulos:2009rq}. Within GENIE (3.0.6), another FSI model, the INC $h$N is available, which features a full cascade approach. To probe the impact of various FSI models on the predictions, Model-G1/3 employs the $h$A model, whereas Model-G2/4 uses the $h$N model for comparison. NuWro (17.10) models adopt the default FSI model~\cite{Golan:2012wx}, which is described in a framework of the INC model~\cite{Metropolis:1958sb}. The specific dynamics is taken from the Oset model~\cite{Salcedo:1987md}. This NuWro FSI model is compared to the NC $\pi^{0}$ production data. NuWro (19.02) has upgraded the FSI model, in which the major improvements include an effective nucleon density considering nucleon-nucleon correlations, implementation of elastic nucleon-nucleon scattering angular coefficients, new inelastic nucleon scattering cross sections, single pion production fractions, and corrections to in-medium inelastic nucleon scattering cross sections~\cite{Niewczas:2019fro}.
\end{description}

In addition to the QE and $2p2h$ interactions, all the other interaction processes such as RES, COH and DIS are also integrated into our calculations, in which the generator default settings are applied to all these processes. For instance, the Rein-Sehgal model~\cite{Rein:1980wg,Rein:1982pf} is applied in Model-G1 to describe RES and COH, while a model using modifications suggested by the Bodek-Yang model~\cite{Bodek:2002ps} is utilized for DIS. The same cross section models of RES, COH and DIS are used in Model-G2/3/4, with the Berger-Sehgal model~\cite{Berger:2008xs} for resonance pion production.

\subsection{TALYS-based Deexcitation Model of Residual Nucleus}
The nuclear deexcitation involves additional $\gamma$ rays, protons, neutrons, or other heavier projectiles. The production of neutrons and unstable residual nuclei during the deexcitation stage is crucial for tagging and reducing potential NC background in LS detectors.
However, Monte Carlo neutrino generators typically lack information on the deexcitation of residual nuclei that may remain excited. 
The implementation of the deexcitation of the residual nucleus from NC interactions between atmospheric neutrinos and $^{12}$C follows the methodology outlined in the preceding paper and must carried out separately from the neutrino generation.

The nuclear structure is specified using a simplified statistical configuration from the nuclear shell model of $^{12}$C~\cite{Kamyshkov:2002wp,Auerbach:1997ay,Kolbe:1999au}, as depicted in Fig.~\ref{fig:ncprocess}.
Additionally, we neglect the potential configuration of the $^{12}$C nucleus arising from nucleon pairing correlation between the $p_{3/2}$ and $p_{1/2}$ shells, given the relatively small energy gap of around 3-4 MeV\cite{Kamyshkov:2002wp,Auerbach:1997ay}. This effect will be included in our future work, incorporating more sophisticated shell model calculation.  
The excited states of residual nuclei are determined by considering the disappearance of one or more nucleons (either protons or neutrons) from the $s_{1/2}$ shell of the $^{12}$C ground state. Note that, in the simplified shell model configuration, if one or more nucleons disappear from the $p_{3/2}$ shell, these nucleons do not impact the state of the residual nucleus. The cases of one or two nucleons disappearing from $^{12}$C in shown in Table II of the preceding paper. This table includes the corresponding probabilities of possible configurations together with the excitation energies of these residual nuclei. There are typos for the configuration probabilities for $^{10}$B$^{*}$ and the corrected information is 4/9, 4/9 and 1/9 for $E^{*} \, = \,0 \, \rm MeV \, (\rm  \, ground \, state), \,23\, \rm MeV, \, 46$ MeV, respectively. 

After determining the excited state of the residual nucleus with the excitation energy $E^{*}$, TALYS (1.8) utilizes this information to simulate all potential deexcitation channels for the nucleus and obtain their deexcitation along with the energy spectra of the associated final-state particles. In our calculations, all residual nuclei with a mass number larger than five have been considered.
For the dominant residual nucleus, $^{11}$C, $^{11}$B, and $^{10}$B, when they remain at $E^{*} = $ 23 MeV, approximately 10\%, 70\%, and 20\% of the deexcitation channels involve neutron production, respectively.
The impact of deexcitation on NC background predictions will be discussed later.

\subsection{GEANT4-based Detector Simulation}

GEANT4 (version 4.10.p02) is utilized to simulate the propagation of final-state particles resulting from $\nu_{\rm atm}$-$^{12}$C NC interactions within the LS detector. The simulation employs the QGSP$\_$BERT physics list, with the Quark Gluon String (QGS) model applied for hadronic interactions at higher energies. For hadronic interactions at the lowest energy range (below 70 MeV) and nuclear deexcitation coupled with the higher-energy QGS model, GEANT4$^{\prime}$s precompound (P) model is employed. Additionally, the Bertini Cascade (BERT) model is applied for hadronic interactions at the lower energy range. The High-Precision neutron (HP) model is used for neutron elastic and inelastic interactions below 20 MeV, ensuring high accuracy in modeling these interactions. The detector simulation incorporates the decay processes of unstable residual nuclei, utilizing information such as decay types, endpoints, and lifetimes sourced from the nuclear database~\cite{database}.

To be specific, we use the JUNO detector~\cite{JUNO:2021vlw} as our reference. For simplicity, the detector simulation excludes the optical process, and the quenching effect in the LS is accounted for using Birk's equation described in Ref.~\cite{An:2015jdp}. Consequently, the detector simulation converts the kinetic energies of final-state particles from NC interactions into quenching energy, representing the visible energy. Relevant time and vertex information is recorded, allowing the selection of events of interest based on associated information, such as single signals and coincident events. This work is specifically dedicated to investigating NC backgrounds within the visible energy range below 100 MeV.

\section{Characteristics of $\nu_{\rm atm}$-$^{12}$C NC Interactions}\label{sec:NCchar}

Using the $\nu_{\rm atm}$-$^{12}$C NC interaction prediction method detailed in Sec.~\ref{sec:Calcalation} and adopting representative models from Tab.~\ref{table:models}, we have simulated the $\nu_{\rm atm}$-$^{12}$C NC interaction in the LS detector. This involves the simulation of secondary interactions caused by final-state particles. The section delves into the characteristics of these interactions across simulation stages, covering $\nu_{\rm atm}$-nucleon NC interactions, FSI, deexcitation of the residual nucleus, and secondary interactions of final-state particles in the detector. We analyze the model variations concerning the $\nu_{\rm atm}$-nucleon NC interactions and final-state information, investigating the impact of FSI, nuclei deexcitation, and secondary interactions on the final-state information.

\subsection{$\nu_{\rm atm}$-nucleon NC Interactions}

First, let us consider the $\nu_{\rm atm}$-nucleon NC interactions, which are fundamental to $\nu_{\rm atm}$-$^{12}$C NC interactions. Understanding these interactions is crucial for comprehending the dynamics of neutrinos and their impact on nucleons within $^{12}$C nuclei. This understanding directly influences the probability of NC interaction between neutrinos and $^{12}$C, determining corresponding event rates. Due to a lack of data constraints, there is a notable level of uncertainty associated with this process. Representative models in Tab.~\ref{table:models} aid in understanding average cross-sections between $\nu_{\rm atm}$ and nucleons within $^{12}$C, as well as the event rate distributions of $\nu_{\rm atm}$-$^{12}$C NC interactions. To understand the model variations of these models in low and high initial neutrino energy ($E_{\nu}$) ranges, we divide the NC interactions into two groups:
\begin{itemize}
    \item Group I: dominant in $E_{\nu} <$ 1 GeV range, includes QE and $2p2h$ interactions with predominantly hadron-only final-state particles.
    \item Group II: dominant in $E_{\nu} >$ 1 GeV range, includes RES, COH, and DIS with predominantly pion production in the final-state particles.
\end{itemize}
  
\subsubsection{Cross-sections}

\begin{figure}[!tb]
    \centering
	\includegraphics[width=0.7\linewidth]{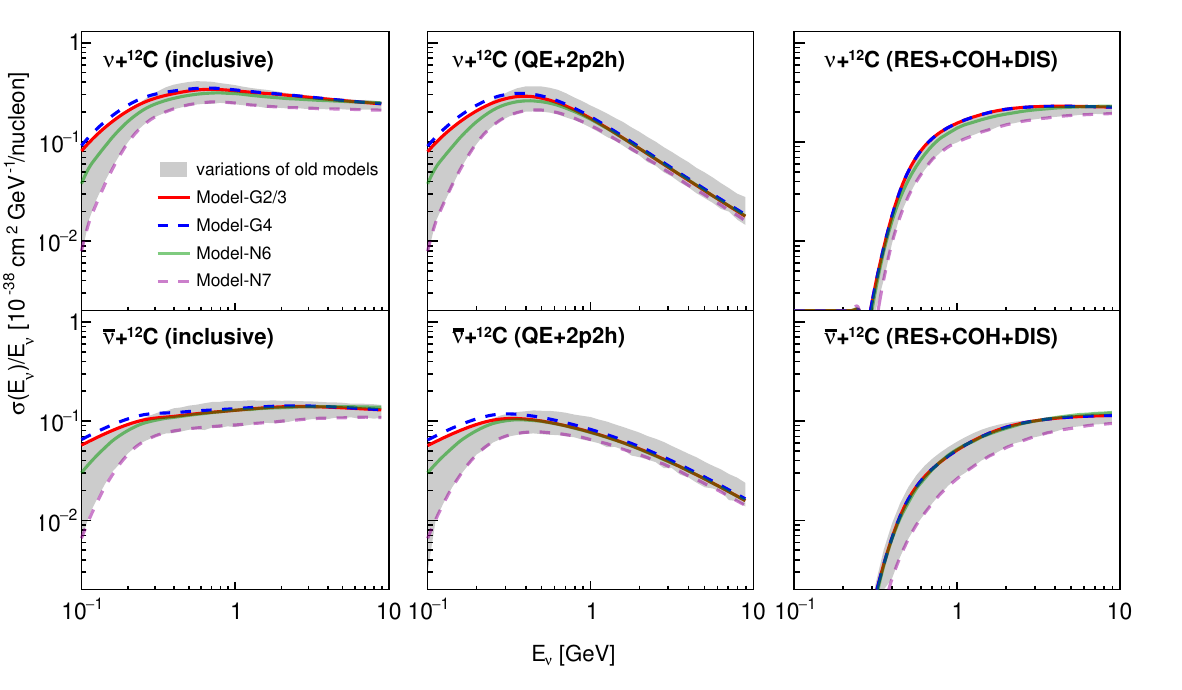}
    \caption{(Anti)neutrinos per nucleon cross-sections $\sigma(E_{\nu})/E_{\nu} \, [10^{-38} \, \rm cm^{2} \, \rm GeV^{-1}/nucleon]$ for the NC interactions and $^{12}$C target are depicted in three panels (from left to right). These panels respectively showcase the inclusive cross-section, the cross-section involving QE and $2p2h$ interactions, and the cross-section involving RES, COH, and DIS processes.}
	\label{fig:ncxsec}      
\end{figure}

Fig.~\ref{fig:ncxsec} illustrates the average cross-sections between $\nu_{\rm atm}$ and nucleons within $^{12}$C, based on the representative models listed in Tab.~\ref{table:models}. The top and bottom panels depict the cross-sections for neutrinos and anti-neutrinos, respectively. The three panels, from left to right, represent the inclusive cross-sections, exclusive cross-sections in group I, and exclusive cross-sections in group II, respectively. In all panels, shaded bands show the variations of $\sigma(E_{\nu})/E_{\nu}$ across the range of values from old models, as a function of the initial neutrino energy.
The smallest cross-section values in all panels are obtained from NuWro, Model-N5 ($M_{\rm A} = 0.99$ GeV for QE and SF). On the other hand, the maximum values predominantly come from NuWro, Model-N3/4. Model-N3 employs $M_{\rm A} = 1.35$ GeV for the QE process, and Model-N4 uses the TEM model to predict the contribution of the $2p2h$ process, resulting in a larger cross-section below 1 GeV, but the TEM effect is absent in the antineutrino sector. Note that in group II, except for Model-N5, the other old models show quite similar cross-section results.

The cross-section results from the new models added in this work are depicted using lines with distinct colors and styles.   
Compared to Model-N3/4, these new models exhibit lower cross-sections.  Note that the cross-sections concern the initial neutrino interactions so that the cross-sections between Model-G2 and Model-G3 are the same. Model-N7 (SF) still offers the smallest value of the inclusive cross-section, consistent with Model-N5 using a different NuWro version. 

In this work, we specifically investigate model variations for the cross-sections in the low energy range, particularly focusing on the QE and $2p2h$ interactions. 
In addition to the insights gained from variations in inclusive cross-sections, we also observed significant differences when examining the specificities of cross-sections in QE and $2p2h$ interactions, as depicted in the middle panel of Fig.~\ref{fig:ncxsec}. These insights are based on the new models used for illustration.
Firstly, the nuclear model significantly influences the cross-section at low energies. This impact can be discerned by comparing models with identical settings but different nuclear models. For instance, a comparison between Model-G2 (LFG) and Model-G4 (BRRFG), or between Model-N6 (LFG) and Model-N7 (SF), clearly illustrates this nuclear model effect. 
Secondly, consistent alignment of cross-sections from Model-G2 and Model-N6 takes place above 300 MeV. However, significant deviations are observed at lower energies. 
This discrepancy is likely attributed to different treatments of the initial nuclear state in GENIE and NuWro.
Thirdly, precise model descriptions of $2p2h$ processes are essential for accurate predictions of atmospheric neutrino NC interactions at low energies. These variations of cross-sections regarding the initial interactions from all representative models are considered as a systematic uncertainty in our calculations.

\subsubsection{$\nu_{\rm atm}$-$^{12}$C NC interaction rates}

\begin{figure}[!tb]
    \centering
	\includegraphics[width=0.7\linewidth]{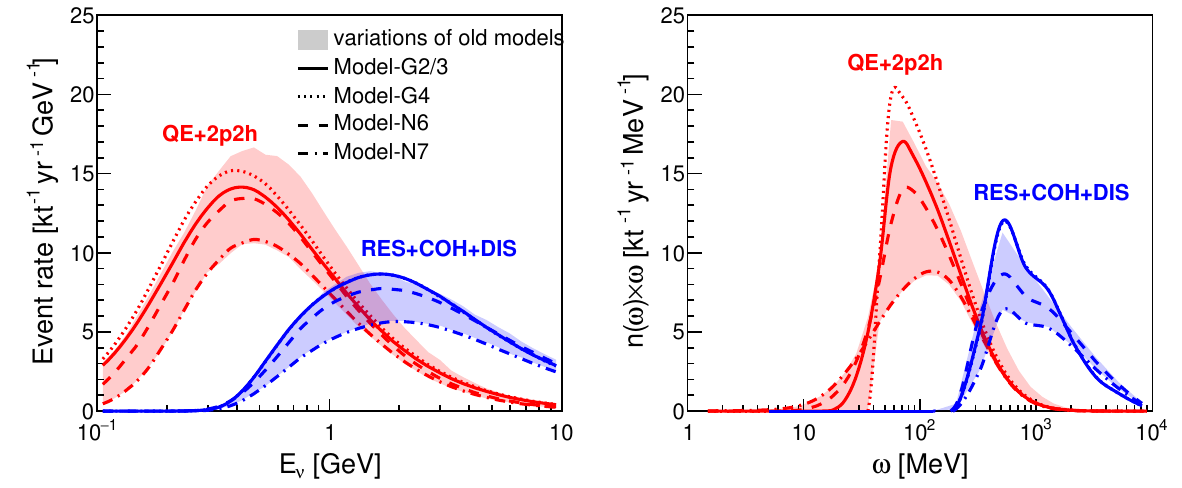}
    \caption{Event rates of two groups of physics processes for NC interactions between neutrinos and $^{12}$C nucleus as a function of incoming neutrino energy (left panel) and energy transfer (right panel). The first group, including QE and $2p2h$, is represented in red, while the second group, including RES, COH and DIS processes, is represented in blue. The shaded bands show the rate variations from the old models used in the preceding paper.}
	\label{fig:ncevandet}      
\end{figure}
 
We utilize the atmospheric (anti)neutrino fluxes at the JUNO site and the cross-section distributions from Fig.~\ref{fig:ncxsec} to calculate the event rate of $\nu_{\rm atm}$-$^{12}$C NC interactions. The relevant energy range of the atmospheric neutrinos used in this work is 100 MeV $< \, E_{\nu}\, <$ 10 GeV, and the total number of the target $^{12}$C nuclei per kiloton LS is around $4.4\times 10^{31}$. The event rates of group I and group II of the atmospheric neutrinos with $^{12}$C nuclei are presented in Fig.~\ref{fig:ncevandet} by summing up all the neutrino species. The rates are depicted concerning the initial neutrino energy and the energy transfer, shown in the left and right panels, respectively. The energy transfer ($\omega$) is defined as the energy difference between incoming and outgoing neutrinos. 

The individual contributions from various physical processes (i.e., QE, $2p2h$, RES, and DIS) to inclusive event rates in representative models (refer to Tab.~\ref{table:models}) have been examined. In group I, consisting of QE and $2p2h$, these interactions collectively contribute approximately 70\% to the total NC event rate within the initial neutrino energy range of 100 MeV to 10 GeV. Given the energies of our interest in this work are relatively low, the QE process is of crucial importance. 
In experimental searches for IBD signals below 100 MeV visible energy, the QE and $2p2h$ processes account for nearly 99\% of the NC backgrounds.
Therefore, to be clear, the following final-state investigations of the NC interaction only focus on QE and $2p2h$.

In Fig.~\ref{fig:ncevandet}, new model variations are depicted alongside old model variations represented by a shaded band. In group I, variations in old models primarily stem from the values of $M_{\rm A}$ and nuclear models. Given that we maintain $M_{\rm A}$ around 1 GeV for new models, variations in low energies are predominantly due to changes in nuclear models. Notably, the BRRFG nuclear model results in the highest QE event rate, while the SF nuclear model corresponds to the lowest QE event rate.
Examining the differential event rates of QE and  $2p2h$  processes for NC interactions in the right panel of Fig.~\ref{fig:ncevandet}, maximum rates occur at $\omega$ = 100 MeV, with the greatest discrepancies among models. The energy transfer distribution's main features are reflected in the final visible energy spectra of NC events in the LS. Consequently, significant variations are expected in the NC background's visible energy spectra below 100 MeV.

\subsection{Final-state Information}
Then, proceeding to the examination of final-state particles in NC events becomes crucial for effective tagging. In $\nu_{\rm atm}$-nucleon NC interactions, neutrinos depart from the nuclei, transferring energy to nucleons as depicted in the right panel of Fig.~\ref{fig:ncevandet}. In the scenario where nucleons are free, the final-state of the QE process comprises solely single nucleon productions. However, these nucleons are bound within the $^{12}$C nuclei. Subsequent processes such as FSI and nucleus deexcitation have the potential to expel additional nucleons, resulting in the emission of multiple final-state particles, including $n$, $p$, $\gamma$, $\alpha$ and others.  
To investigate the influence of FSI and deexcitation on final-state particle production, we employ new models to analyze and compare their effects across different FSI models, evaluating outcomes when deexcitation is either activated or deactivated.
  
\subsubsection{Production rates of final-state particles}

\begin{figure}[!tb]
    \centering
	\includegraphics[width=0.5\linewidth]{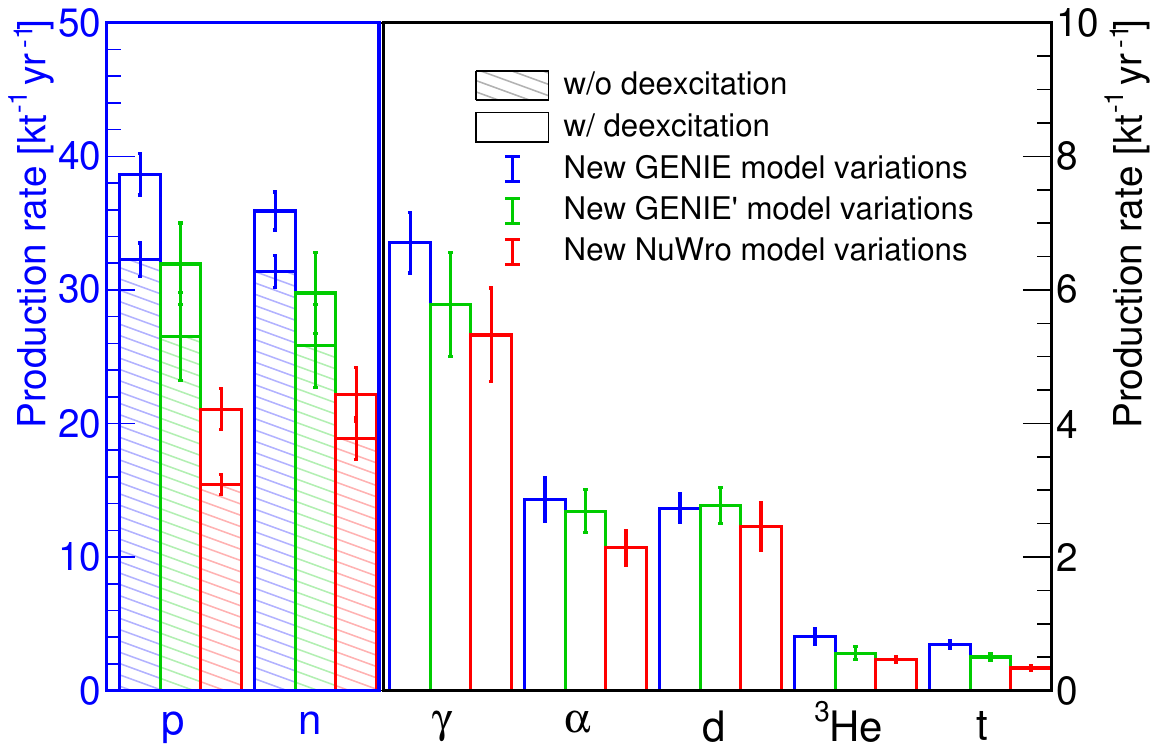}
    \caption{The model variations in production rates of final-state particles from both QE and $2p2h$. The median rates without and with deexcitation are shown as the striped histogram with backslashes and hollow-filled histogram, respectively. The error bar represents the model variations in production rates. The production rates of $n$ and $p$ are represented on the left y-axis (blue), while other entities are aligned with the right y-axis (black).}
	\label{fig:produrate}      
\end{figure}

Initially, we calculate the production rates of relevant final-state particles from both QE and $2p2h$. However, observation reveals that GENIE (3.0.6) generates nucleons with zero kinetic energy, a phenomenon also identified in the older GENIE model~\cite{Cheng:2020oko}. Our investigation extends to the newer version of GENIE (3.4.0), where the issue of final-state nucleons with zero kinetic energies persists. Consultation with GENIE's authors clarifies that this problem stems from the handling of binding energy during the NCQE process in GENIE. Note that the most recent GENIE versions are anticipated to improve the handling of binding energy in the modeling of the FSI processes. To address the issue, their recommendation involves marking a nucleon as intermediate if it lacks sufficient kinetic energy to leave a nucleus (e.g., its kinetic energy inside the nucleus is lower than the binding energy), with all its energy-momentum absorbed by the hadronic blob. Following these suggestions, the final-state information from the new GENIE models (i.e., Model-G$i$ for $i = $ 2, 3, 4) is modified. To distinguish models subjected to this operation from those that are not, the modified models are re-labeled as Model-G$i$ for $i = $ 2$^{\prime}$, 3$^{\prime}$, 4$^{\prime}$. 

Analyzing final-state results from all new models reveals significant distinctions in the comparison between GENIE and NuWro models. In contrast, GENIE models exhibit similarity, and NuWro models also show similarity. To illustrate model variations among the new models, we categorize them based on associated neutrino generators and divide them into three categories:
\begin{itemize}
    \item ``New GENIE model variations": the model variations among Model-G$i$ for $i = $ 2, 3, 4.
    \item ``New GENIE$^{\prime}$ model variations": the model variations among Model-G$i$ for $i = $ 2$^\prime$, 3$^{\prime}$, 4$^{\prime}$, which are used to demonstrate the impact of modifications in the new GENIE models, particularly addressing the issue of zero kinetic energy nucleons production in final-state results.
    \item ``New NuWro model variations": the model variations between Model-N6 and Model-N7. 
\end{itemize}

In Fig.~\ref{fig:produrate}, we provide a summary of model variations for the above three categories in the production rate of final-state particles from QE and $2p2h$. Results excluding and including the deexcitation of the residual nucleus are shown as the striped histogram with backslashes and hollow-filled histogram, respectively. When focusing on the final-state particle productions before deexcitation, the primary final-state particles are protons and neutrons. Note that the contribution of $\pi$'s produced by FSI processes to final-state particles is very small and can be considered negligible. Without deexcitation, each new GENIE model exhibits comparable production rates for protons and neutrons, while new NuWro models yield a higher proportion of neutrons in comparison to protons. The upper and lower limits of the error bars indicate model variations. Since neutron products can be identified by the LS detector, some comments for the model variations of production rates of neutrons excluding these induced by deexcitations are helpful. 

\begin{itemize}
 \item In the new GENIE models, Model-G2 exhibits the lowest rates, while Model-G4 shows the highest rates. The distinction between these models lies in the choice of both nuclear and FSI models. Specifically, Model-G4 (BRRFG), produces around 8\% higher rate of neutrons compared to Model-G2 (LFG). Additionally, Model-G3 ($h$A), yields about 6\% higher rates of neutrons compared to Model-G2 ($h$N).

 \item When excluding nucleons with zero kinetic energies in the final-state, Model-G2$^{\prime}$/4$^{\prime}$ ($h$N) exhibit a 25\% lower rate of neutrons compared to the associated models. Similarly, Model-G3$^{\prime}$ ($h$A) shows a 10\% lower rate of neutrons in comparison. Therefore, in the new GENIE$^{\prime}$ models, Model-G2$^{\prime}$ still exhibits the lowest rates, while Model-G3$^{\prime}$ shows the highest rates.

 \item In the new NuWro model variations, Model-N6 (LFG) generates $\sim$18\% higher neutron rates compared to Model-N7 (SF). Furthermore, NuWro models exhibit significantly lower neutron rates than those from GENIE models. These differences may be subject to constraints imposed by future measurements of $\nu_{\rm atm}$-$^{12}$C NC interactions using large LS detectors.
\end{itemize}

When comparing production rates with and without deexcitation, neutron production rates, including deexcitation, show an enhancement of around (13 $-$ 19)\% compared to those without deexcitation for all new models. Additionally, deexcitation induces almost all $\gamma$'s, $\alpha$'s, $d$'s, $^{3}$He's, and $t$'s particles. Note that variations in the models of products from deexcitation are associated with the model variations of the excited residual nucleus.

\begin{figure}[!tb]
    \centering
	\includegraphics[width=0.7\linewidth]{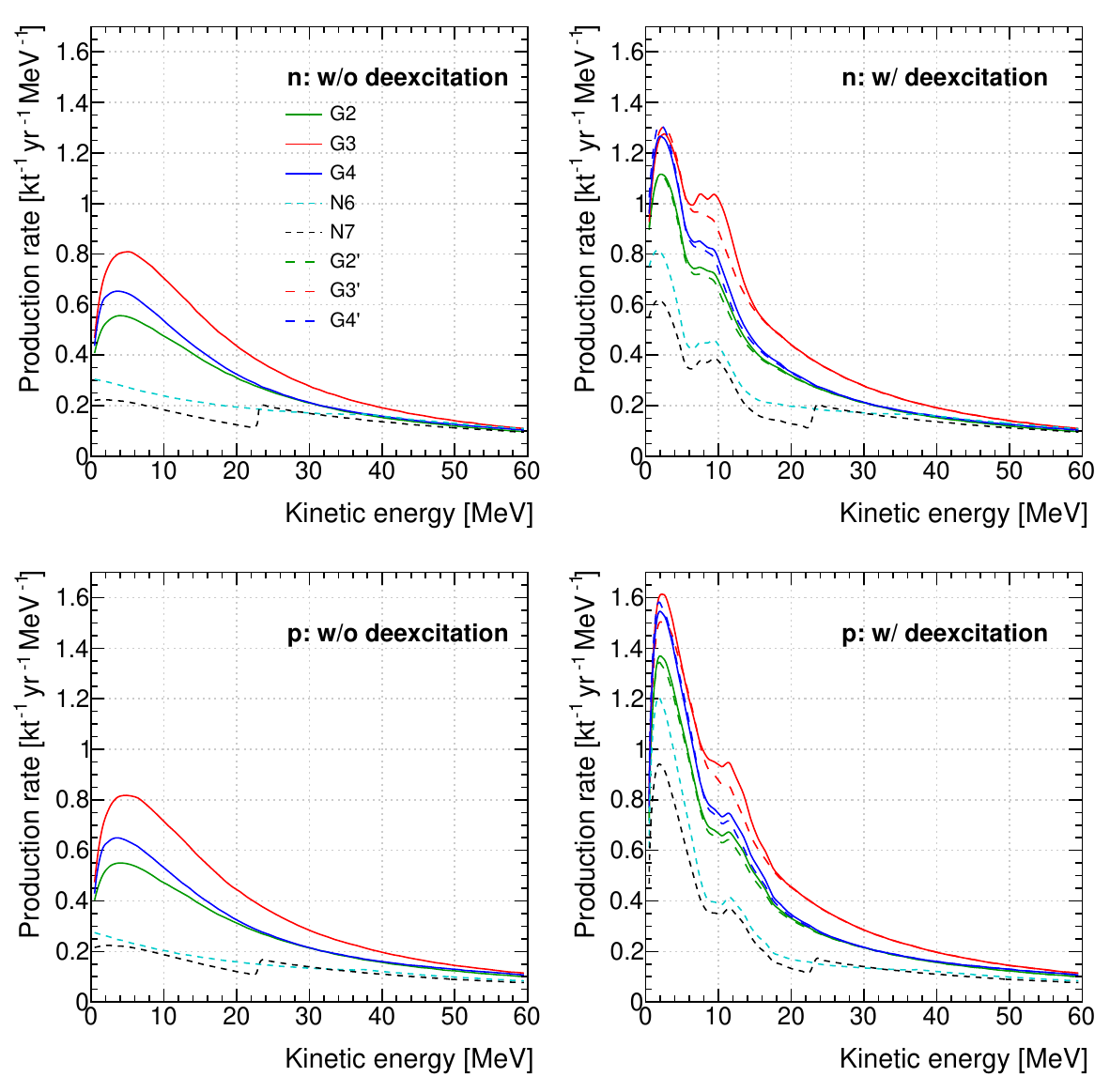}
    \caption{The kinetic energy spectra of $n$ (upper panels) or $p$ (low panels) from the QE and $2p2h$ interactions of $\nu_{\rm atm}$-$^{12}$C from new models. The left panels show the spectra of $n$ and $p$ from interactions without deexcitation, while the right panels depict the spectra with deexcitation. Nucleons with zero kinetic energy are omitted for clarity.}
	\label{fig:ek}      
\end{figure}

Fig.~\ref{fig:ek} presents the production rates of primary final-state particles, such as $n$ and $p$, for kinetic energy. The dominant contributors to production are the QE and $2p2h$. To enhance clarity, nucleons with zero kinetic energy are excluded in Fig.~\ref{fig:ek}. This operation results in kinetic energy spectra in Model-G$i =$ 2, 3, 4 aligned with those from the corresponding Model-G$i =$ 2$^{\prime}$, 3$^{\prime}$, 4$^{\prime}$ in the left panels of Fig.~\ref{fig:ek}, illustrating the spectra of $n$ and $p$ from NC interactions without deexcitation. When focusing on nucleon productions excluding those from deexcitation, we observe the same conclusions as depicted in Fig.~\ref{fig:produrate}. Furthermore, the impact of different nuclear models on production rates falls within the (0 $-$ 20) MeV range in kinetic energy, as evident in the comparison between Model-G2 and Model-G4, or Model-N6 and Model-N7. Similarly, the effects of different FSI models on production rates are within the (0 $-$ 40) MeV range in kinetic energy, observed in comparisons such as Model-G2 versus Model-G3, or Model-G2 versus Model-N6. By comparing the left and right panels of Fig.~\ref{fig:ek}, it becomes apparent that the enhanced production rates of $n$ and $p$ from deexcitation occur within the (0 $-$ 20) MeV range of kinetic energy. As expected, due to slight differences in the excited residual nucleus, the kinetic energy spectra of Model-G$i =$ 2$^{\prime}$, 3$^{\prime}$, 4$^{\prime}$ exhibit minor variations from the corresponding Model-G$i =$ 2, 3, 4.

\subsubsection{Neutron multiplicity}

\begin{figure}[!tb]
    \centering
	\includegraphics[width=0.7\linewidth]{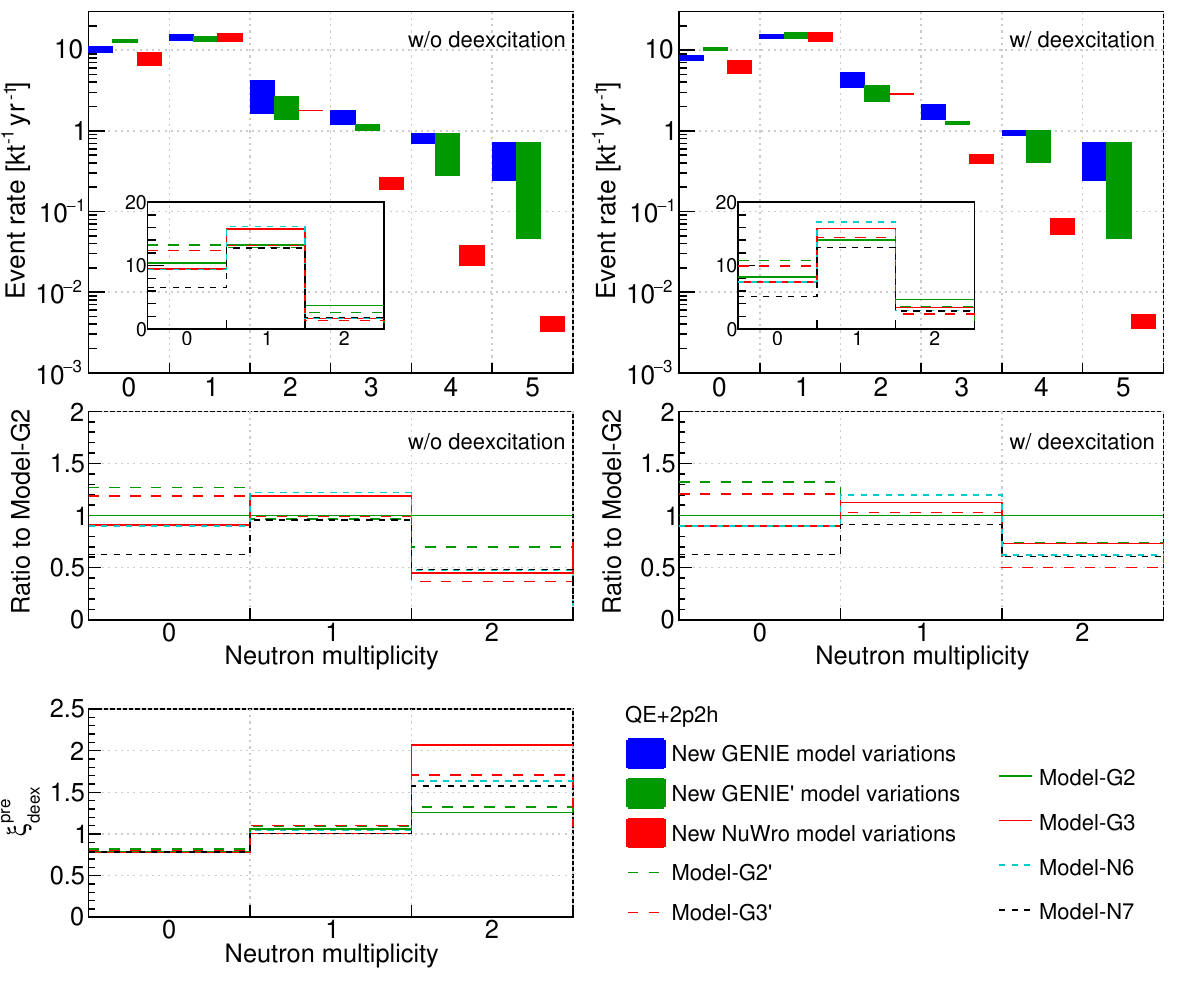}
    \caption{The event rates for NC interactions of $\nu_{\rm atm}$-$^{12}$C as a function of neutron multiplicity. In the top panels, the event rates for neutron multiplicities less than three are specifically displayed in the subpanels. The middle panels depict the ratio of the event rate in each new model to that in Model-G2, while the bottom panels illustrate the ratio ($\xi_{deex}^{pre}$) of the event rate with deexcitation to that without deexcitation.}
	\label{fig:nmbfdet}      
\end{figure}

\begin{figure}[!tb]
    \centering
	\includegraphics[width=0.7\linewidth]{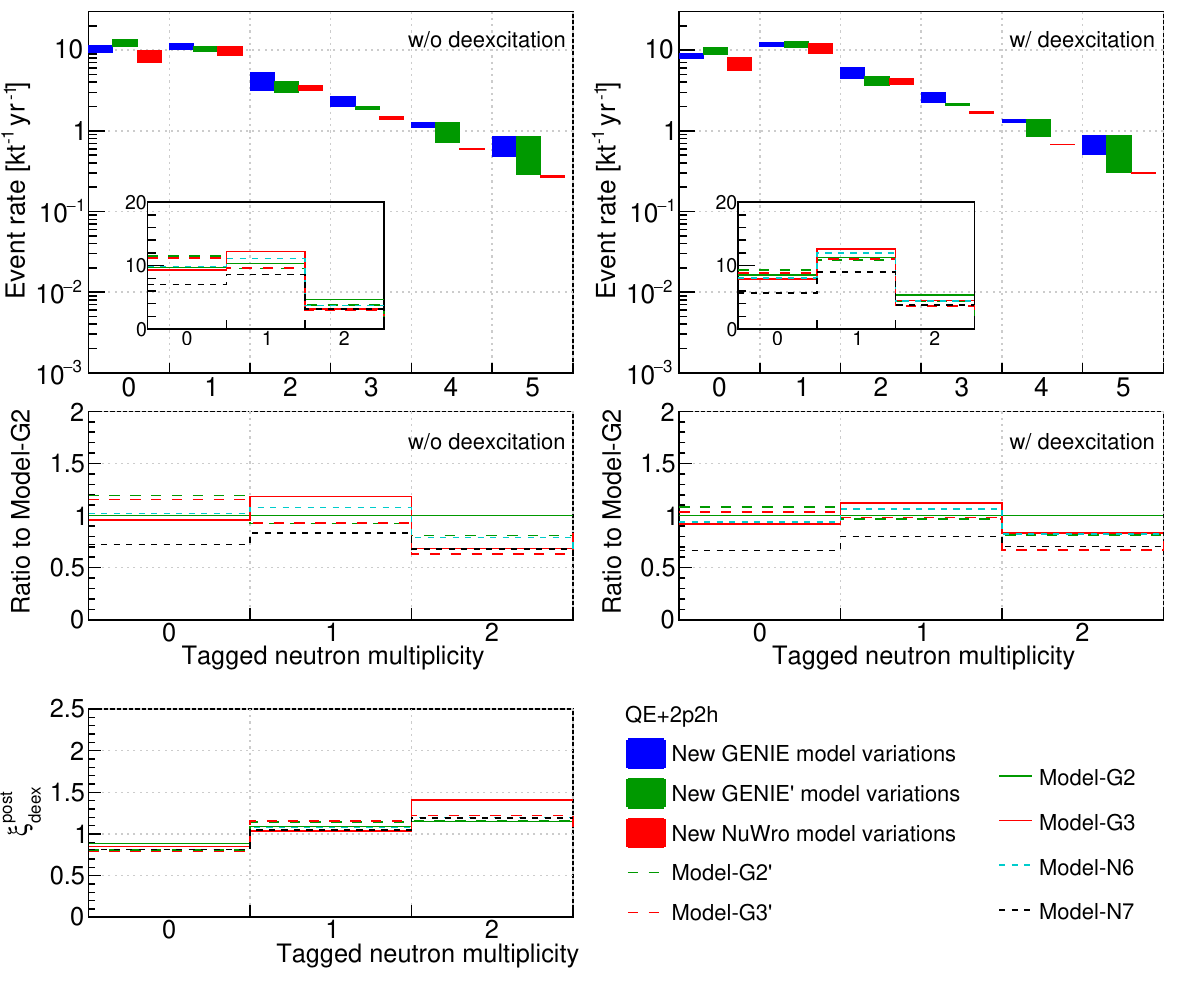}
    \caption{The event rates for NC interactions of $\nu_{\rm atm}$-$^{12}$C as a function of tagged neutron multiplicity in the LS. In the top panels, the event rates for neutron multiplicities less than three are specifically displayed in the subpanels. The middle panels depict the ratio of the event rate in each new model to that in Model-G2, while the bottom panels illustrate the ratio ($\xi_{deex}^{post}$) of the event rate with deexcitation to that without deexcitation.}
	\label{fig:nmafdet}      
\end{figure}

Neutrons are efficiently tagged, through their captures on H or C atoms in the LS detectors. Neutron tagging plays a crucial role in investigating $\nu_{\rm atm}$ NC events. Consequently, the event rates of the QE and $2p2h$ are presented in Fig.~\ref{fig:nmbfdet} and Fig.~\ref{fig:nmafdet} based on the neutron multiplicity. The vivid bands depict variations among the new models in the three mentioned categories.  

In Fig.~\ref{fig:nmbfdet}, we examine the neutron multiplicity in the final-state of NC interactions involving $\nu_{\rm atm}$-$^{12}$C. Considering that the predominant neutron multiplicities are below three, the event rates for neutron multiplicities less than three are highlighted in the subpanels of the upper panels of Fig.~\ref{fig:nmbfdet}. Due to the significant similarity in distributions between Model-G2 and Model-G4, as well as between Model-G2$^{\prime}$ and Model-G4$^{\prime}$, Model-G4/4$^{\prime}$ is excluded from the illustrations. Excluding neutrons from the deexcitation process, the events with neutron multiplicities less than three account for approximately 91\% in the new GENIE models using the $h$N FSI model (i.e., Model-G2/4) and $h$A FSI model (i.e., Model-G3). This calculation includes neutrons with zero kinetic energies in the final-state.
If we exclude these neutrons in Model-G$i =$ 2$^{\prime}$, 3$^{\prime}$, 4$^{\prime}$, the events with neutron multiplicities less than three increase to 95\% in Model-G2$^{\prime}$/4$^{\prime}$, while the fraction remains unchanged in Model-G3$^{\prime}$. This difference arises because neutrons with zero kinetic energies from the $h$N FSI model impact all neutron multiplicities, whereas those from the $h$A FSI model only affect neutron multiplicities less than three. Compared to GENIE models, NuWro models produce around 99\% events with neutron multiplicities less than three. For comparison purposes, the middle panels of Fig.~\ref{fig:nmbfdet} display the ratio of results related to multiplicities less than three to those in Model-G2. The neutrons induced by the deexcitation process increase the multiplicity. The parameter, $\xi_{deex}^{pre}$, defined as the ratio of event rates for NC events with deexcitation to those for NC events without deexcitation, concerning multiplicities less than three is illustrated in the lowest panel of Fig.~\ref{fig:nmbfdet}.  

Fig.~\ref{fig:nmafdet} presents a summary of the final event rates for QE and $2p2h$ interactions of $\nu_{\rm atm}$-$^{12}$C as a function of tagged neutron multiplicity, where tagged neutron multiplicity is defined as the numbers of neutrons captured on H or C atoms in the LS after an NC interaction. The notations and patterns of the histograms precisely mirror those in Fig.~\ref{fig:nmbfdet}. Specifically, the parameter $\xi_{deex}^{post}$ is defined as the ratio of event rates for NC events with the deexcitation process to those for NC without the deexcitation process, relative to tagged neutron multiplicities less than three.
We expect that the event rates linked to tagged neutron multiplicities in the LS detector should align with the event rates of produced neutron multiplicities as depicted in Fig.~\ref{fig:nmbfdet}. 
However, what we observe are those after secondary interactions in LS. 
For instance, the energetic fast neutrons from the $\nu_{\rm atm}$-$^{12}$C NC interactions may disappear or produce other neutrons due to their inelastic interactions with $^{12}$C. This effect has been investigated by using TALYS to calculate the cross-sections of the exclusive reaction channels at different incident neutron energies~\cite{Cheng:2020oko,Cheng:2023zds}. In the case of fast neutrons with kinetic energies below 10 MeV, the primary interaction between neutrons and $^{12}$C is elastic scattering ($n^{12}$C). However, as the energy surpasses 10 MeV, the inelastic scattering cross-section of $n^{12}$C experiences a substantial increase, leading to competition with the elastic scattering process. Comparing Fig.\ref{fig:nmbfdet} and Fig.\ref{fig:nmafdet}, it is evident that the event rates for neutron multiplicities equal to one decrease significantly due to the secondary interaction effects. Furthermore, the differences between models are diminished as a result of the secondary interaction effects.

\begin{figure}[!tb]
    \centering
	\includegraphics[width=0.7\linewidth]{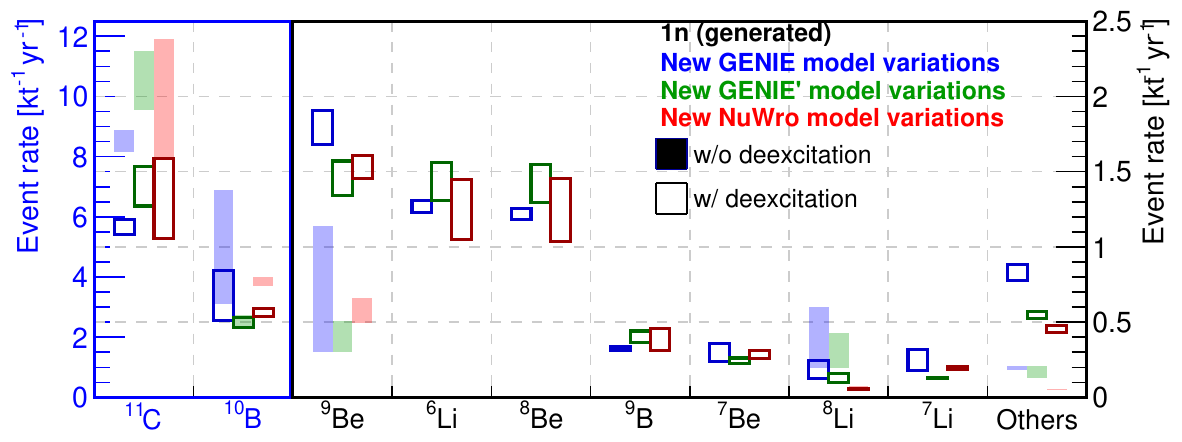}
 \includegraphics[width=0.7\linewidth]{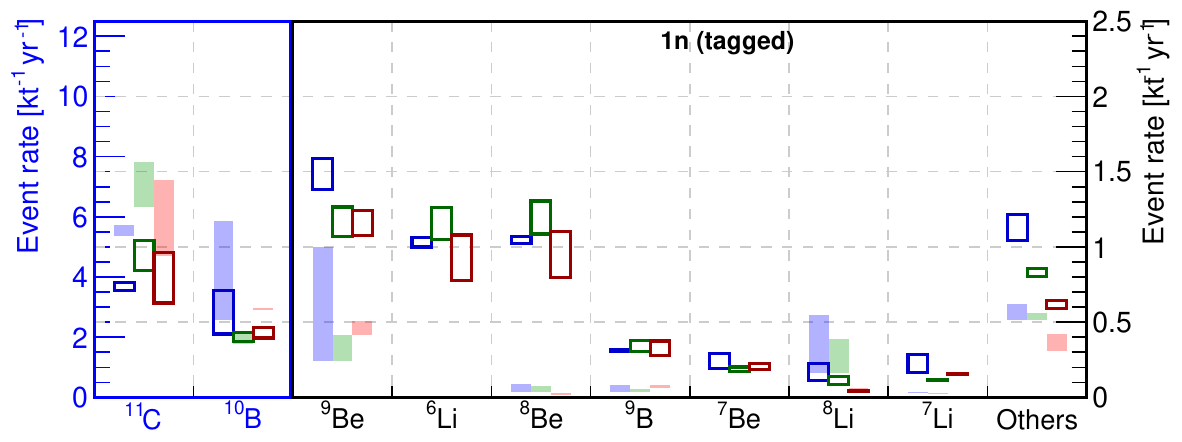}
    \caption{The event rates for $\nu_{\rm atm}$-$^{12}$C QE and $2p2h$ interactions in exclusive channels, focusing on cases where the generated neutron multiplicity is one in the final-state of the NC interaction (top panel) and the tagged neutron multiplicity is one following an NC interaction in the LS (bottom panel)}.
	\label{fig:1niso}      
\end{figure}

Focusing on cases with only one neutron production in the final-state, the event rate for NC events with a neutron multiplicity equal to one in Fig.~\ref{fig:nmbfdet} is categorized by exclusive channels, as depicted in the top panel of Fig.~\ref{fig:1niso}. This panel highlights the predominant influence of processes leading to the production of $^{11}$C and $^{10}$B in NC events without deexcitation. Notably, there is a significant increase in event rates for $^{9}$Be, $^{6}$Li, $^{8}$B, $^{7}$Be, and $^{7}$Li, as well as the ``others" category in NC events with deexcitation. Simultaneously, event rates associated with $^{11}$C and $^{10}$B processes experience substantial reduction due to the further knockout of one or more nucleons during deexcitation.

Upon scrutinizing model variations, the maximum difference is observed in NC events associated with $^{10}$B for the new GENIE model variations. However, considering that nucleons with zero kinetic energy remain within the nuclei, the disparity in NC events with $^{10}$B is significantly mitigated in new GENIE$^{\prime}$ model variations, resulting in lower event rates. Meanwhile, the event rate for NC events with $^{11}$C sees a noteworthy increase in new GENIE$^{\prime}$ models. Regarding new NuWro model variations, the primary distinction arises in NC events with $^{11}$C, while other channels exhibit similarity in the absence of deexcitation scenarios. Notably, Model-N6 with the LFG nuclear model produces substantially higher event rates for NC events with $^{11}$C.

Similarly, the event rate for NC events with one tagged neutron following the subsequent interactions of product particles in the detector in Fig.~\ref{fig:nmafdet} is categorized by exclusive channels. This is illustrated in the bottom panel of Fig.~\ref{fig:1niso}, revealing the impact of secondary interactions on the exclusive channels.

Upon comparing Fig.~\ref{fig:nmbfdet} and Fig.~\ref{fig:nmafdet}, we observe the maximum difference exists in the event rates for neutron multiplicity equal to one. Further investigation reveals that the NC events with $^{11}$C are significantly decreased under the condition of only one tagged neutron in LS. This is due to the higher kinetic energies of fast neutrons from the NC events with $^{11}$C, making them more prone to inelastic scattering with $^{12}$C. This effect has been illustrated in previous work (e.g., see distributions of the kinetic energies of fast neutrons in different exclusive channels in Ref.~\cite{Cheng:2020oko} and the cross-sections of exclusive $n-^{12}$C reactions in Ref.~\cite{Cheng:2023zds}). While this is advantageous for reducing NC background levels in related topics~\cite{Cheng:2023zds}, it poses a challenge for $in$ $situ$ measurements of NC background based on the triple-coincident signatures of the NC events with $^{11}$C~\cite{Cheng:2020oko}.

\section{NC Background Predictions at Low Visible Energies}\label{sec:res}

In this section, we explore the predictions of the NC background within the visible energy below 100 MeV. More specifically, our focus in this study revolves around examining the signals arising from IBD events and potential backgrounds originating from atmospheric neutrino NC interactions within an LS detector. The final-state particles of the NC interactions, such as $p$, $n$, $\alpha$, and $\gamma$ rays promptly deposit their kinetic energies in the LS. In cases where a neutron capture on H or C atoms follows coincidentally within the time frame and distance region, the resulting prompt scintillation signals can mimic the prompt event of an IBD coincidence.

\begin{figure}[!tb]
    \centering
	\includegraphics[width=0.7\linewidth]{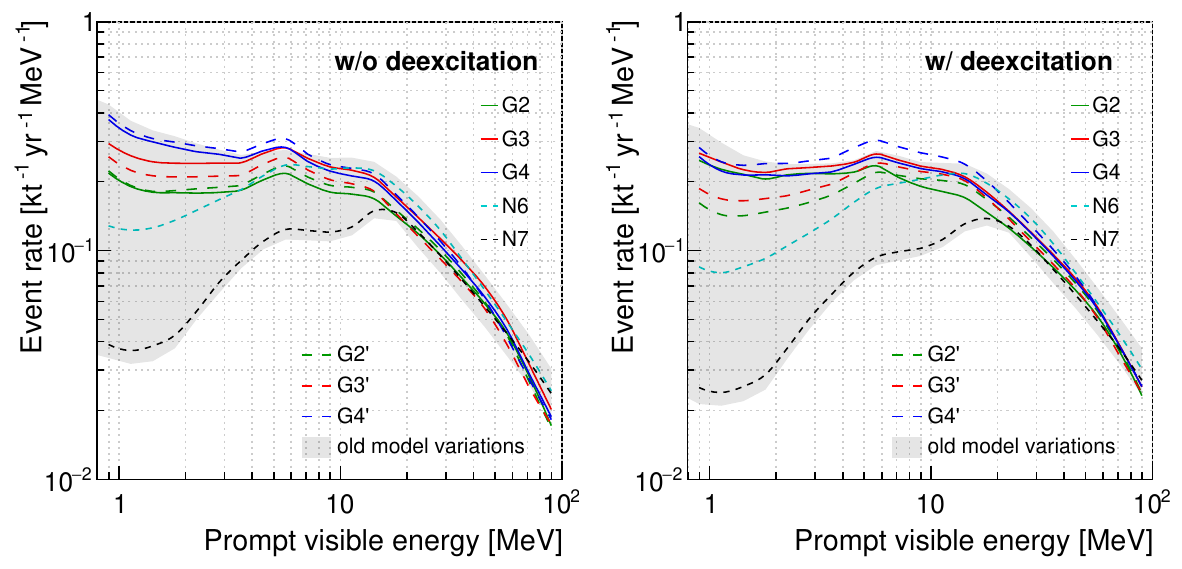}
    \caption{Event rates of the NC background as a function of the prompt energy using all old and new neutrino interaction models.}
	\label{fig:ep}      
\end{figure}

The prompt energy is derived through the GEANT4 detector simulation, encompassing the complete chain of detector response. In Fig.~\ref{fig:ep}, we depict the event rates of the NC background as a function of prompt energy, utilizing both old and new neutrino interaction models to assess the systematic uncertainty of the model prediction. To examine the impact of residual nucleus deexcitation on the final NC background prediction, we present predictions without and with the deexcitation effect in the left and right panels of Fig.~\ref{fig:ep}, respectively. Some comments on Fig.~\ref{fig:ep} are helpful.
\begin{itemize}
    \item Without incorporating the deexcitation process into the NC background prediction, there is minimal variation in both rates and spectra across all neutrino interaction models for energy ranges exceeding 20 MeV. However, a significant disparity is observed within the energy range of (1 $-$ 10) MeV. This discrepancy arises from the model variations of predicting the kinetic energy of final-state particles induced by neutrino interactions, as shown in Fig.~\ref{fig:ek}. For the kinetic energy less than 20 MeV, the model predictions lack sufficient experimental constraints. Hence, predictions in this region require improvement for proper modelling of the detector response.
    When including the deexcitation process in predicting the NC background, the spectra with rate weight, as shown in the right panel of Fig.~\ref{fig:ep}, generally exhibit a shift towards higher energies compared to the case without deexcitation in the left panel. Additionally, the deexcitation process amplifies the model differences for promptly visible energies less than 20 MeV.
    \item In the case of the new GENIE models, note that the rates and spectra exhibit a peculiar similarity for visible energies below 5 MeV in the right panel of Fig.~\ref{fig:ep}. This phenomenon arises due to the inclusion of neutrons with zero kinetic energy in the predictions. The corresponding prompt energies primarily originate from deexcitation productions, and the deexcitation method remains consistent across all neutrino interaction models.
\end{itemize}

\begin{figure}[!tb]
    \centering
	\includegraphics[width=0.7\linewidth]{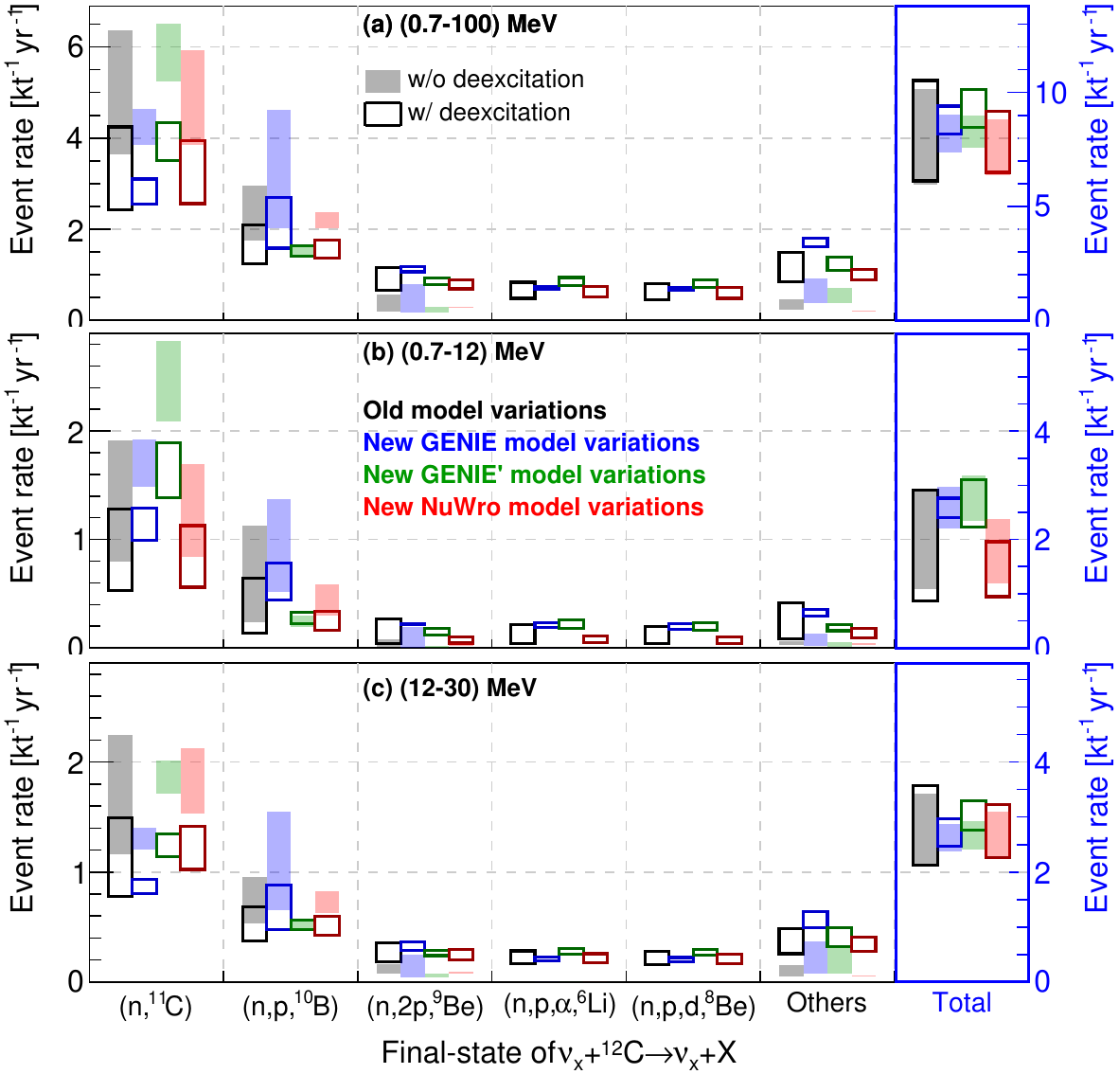}
    \caption{The event rates of NC background in specific channels with various final-state nuclei in the top, middle, and bottom panels for prompt energy ranges of (0.7 $-$ 100) MeV, (0.7 $-$ 12) MeV, and (12 $-$ 30) MeV, respectively.}
	\label{fig:FR}      
\end{figure}

\renewcommand{\arraystretch}{1.2}
\begin{table}[!tb]
\centering
\caption{
The average event rates of the NC background, with the model uncertainty represented by the 1$\sigma$ deviation, for prompt energies in the ranges (0.7 $-$ 100) MeV, (0.7 $-$ 12) MeV, and (12 $-$ 30) MeV. The value from the old models for the prompt energy within (12 $-$ 30) MeV has been adopted by the DSNB study in JUNO~\cite{JUNO:2022lpc}.}
\begin{tabular}{lccc}
\hline \hline
Models & \multicolumn{3}{c}{ Prompt visible energy ranges} \\ 
& (0.7 $-$ 100) MeV   & (0.7 $-$ 12) MeV   & (12 $-$ 30) MeV  \\
\hline
\multicolumn{4}{c}{\underline{NC background rate without deexcitation (kt$\cdot$yr)$^{-1}$}} \\
Old models & 8.4 $\pm$ 1.4  & 2.3 $\pm$ 0.6   & 2.8 $\pm$ 0.5 \\
New models & 7.9 $\pm$ 1.0  & 2.3 $\pm$ 0.6  & 2.6 $\pm$ 0.3  \\

\multicolumn{4}{c}{\underline{NC background rate with deexcitation (kt$\cdot$yr)$^{-1}$}} \\
Old models & 8.9 $\pm$ 1.4 & 2.0 $\pm$ 0.6   & 3.0 $\pm$ 0.5~\cite{JUNO:2022lpc} \\
New models & 8.6 $\pm$ 1.2 & 2.1 $\pm$ 0.7  & 2.9 $\pm$ 0.4  \\
\hline \hline
\end{tabular}
\label{table:ERv2}
\end{table}

Fig.~\ref{fig:FR} presents event rates with different final-state nuclei across distinct prompt energy ranges of (0.7 $-$ 100) MeV, (0.7 $-$ 12) MeV, and (12 $-$ 30) MeV, where the last two energy regions are associated with the detection regions of reactor neutrino and DSNB signals, respectively. The predominance of the NC background with $^{11}$C is evident. The deexcitation effect minimally impacts the overall event rate, causing a reduction in rates for dominant channels with $^{11}$C and $^{10}$B, while increasing rates for channels with lighter nuclei such as $^{9}$Be, $^{6}$Li and $^{8}$Be. When comparing the new and old neutrino interaction models, the rates of the new models do not surpass those of the old models, with only slightly higher rates observed for new GENIE$^{\prime}$ models in the (0.7 $-$ 12) MeV energy range.

Tab.~\ref{table:ERv2} summarizes the average rates of the NC background rates, with the model uncertainty represented by the 1$\sigma$ deviation, for both old and new neutrino interaction models. In this table, the old models considered are Model-G1 and Model-N$i$, where $i=$ 1, 2, 3, 4, 5. Due to a strong correlation between the new GENIE and GENIE$^{\prime}$ models, we opt to use the new GENIE$^{\prime}$ model calculations instead of the new GENIE models for the prediction. Consequently, the results from the new models in Tab.~\ref{table:ERv2} include Model-G$i$, $i=$ 2$^{\prime}$, 3$^{\prime}$, 4$^{\prime}$, and Model-N$i$, $i=$ 6, 7.
To investigate the impact of the deexcitation on the final NC background rate, both results without and with deexcitation are presented in Tab.~\ref{table:ERv2}. It's important to note that while the deexcitation model used in this work still requires constraints from experimental data, the NC background rate with deexcitation is considered as the nominal result, as deexcitation always occurs in a realistic condition.
In the prompt range of (0.7 $-$ 100) MeV, the average NC background rates of the new models are lower than those of the old models for both scenarios, with and without the deexcitation process. This difference arises from the fact that the new models no longer incorporate the model with $M_{\rm A}=$ 1.35 GeV, which previously led to a higher NC background rate in the higher energy range. The deexcitation effect results in an increase in the average NC background rates by around a factor of 1.06 and 1.09 for old models and new models in this energy range, respectively. When considering the deexcitation, the associated uncertainty arising from model variation is around 16\% and 14\% for old and new models, respectively. In the (0.7 $-$ 12) MeV and (12 $-$ 30) MeV energy ranges, we will discuss the results related to the detection of reactor $\bar{\nu}_{e}$ and DSNB in the LS detectors.

\subsection{Reactor $\bar{\nu}_{e}$}

Reactor $\bar{\nu}_{e}$ originates from the $\beta$-decays of neutrino-rich fission fragments, primarily arising from four fission isotopes: $^{235}$U, $^{238}$U, $^{239}$Pu, and $^{241}$Pu. Their detection takes place through the IBD reaction. Taking the JUNO detector as an example, we have calculated the IBD rate and spectrum within the energy range extending from 0.7 MeV to 12 MeV~\cite{JUNO:2022lpc}. The expected IBD signal rate, accounting for the oscillation effect, is 1514.8 yr$^{-1}$kt$^{-1}$.

The NC background spectra and rates for both exclusive channels and the inclusive one can be extracted from Fig.~\ref{fig:ep} and Fig.~\ref{fig:FR}. From Tab.~\ref{table:ERv2}, the NC background rates exhibit a notable similarity between old and new models. This similarity arises from the fact that the primary distinction between old and new models lies in the setting of $M_{\rm A}=$ 1.35 GeV in Model-N3 for the old models, which mainly affects the higher energy range, as mentioned earlier. Additionally, the NC background rates experience a reduction of around 13\% for both old and new models in this visible energy range due to the deexcitation effect. The associated uncertainty, represented by the model variations, ranges from around 26\% to 30\% when comparing rates without deexcitation to rates with deexcitation. The dominant source of uncertainty is attributed to the variation in nuclear effect models. For conservative purposes, the maximum NC background level is derived from the GENIE model, specifically 2.9 kt$^{-1}$ yr$^{-1}$ from Model-G1 in old models and 3.1 kt$^{-1}$ yr$^{-1}$ from Model-G4 in new models.

\subsection{DSNB}

Based on our discussions in prior papers~\cite{Cheng:2020aaw,JUNO:2022lpc}, we have determined the optimized observed prompt energy range for DSNB discovery potential to be from 12 MeV to 30 MeV. In this energy range, the typical predicted rate of DSNB IBD signals in LS detectors is around (0.1 $-$ 0.2) kt$^{-1}$ yr$^{-1}$. The prediction of the NC background within this energy range has been performed using the old neutrino interaction models. By averaging six old model calculations for prediction and considering model variations as uncertainty, we obtained a value of (3.0 $\pm$ 0.5) kt$^{-1}$ yr$^{-1}$ for the NC background in the prompt energy range from 12 to 30 MeV. In this work, we have updated the prediction with new neutrino generator models. The average rates with model variations for both old and new models are shown in Tab.~\ref{table:ERv2}. Since we do not use a higher value of $M_{\rm A}$ in the new models, the average rates become lower, and the model variation becomes smaller in this energy range. Additionally, the associated uncertainty, represented by the model variations, ranges from around 12\% to 14\% when comparing rates without deexcitation to rates with deexcitation.  

Note that the predicted NC background level is about 1 order of magnitude higher than the DSNB signal rate. Therefore, in the pursuit of DSNB discovery, reducing the uncertainty and rate level of the NC background becomes critically important. To address the former, the decays of unstable final-state nuclei may offer unique signatures, allowing for $in$ $situ$ measurements of the NC backgrounds. We have implemented the methodology for $in$ $situ$ measurements~\cite{Cheng:2020oko}. As for the latter, Pulse Shape Discrimination (PSD) provides an effective technique for background suppression of non-positron prompt events~\cite{Cheng:2023zds}. 

\section{Summary}
\label{sec:sum}

In this work, we have extended the methodology of predicting the $\nu_{\rm atm}$ NC interactions 
by boosting its completeness and applicability in the neutrino interactions and the propagation of final-state particles in large LS detectors. Several outstanding advancements are achieved including the incorporation of five new data-driven models from GENIE (3.0.6) and NuWro (19.02) and the secondary interactions in the LS medium. With a focus on deepening our understanding of GeV neutrino interactions, our calculations keep pace with the persistent optimization and validation of neutrino interaction models. 
Moreover, a GEANT4-based detector simulation has been formally integrated into our calculations, paving the way for the examination of secondary interactions of final-state particles. 
The $\nu_{\rm atm}$ NC interactions are significant backgrounds for the experimental searches of the IBD signals of $\bar{\nu}_{e}$ in the visible energy less than 100 MeV, specifically for reactor and geo $\bar{\nu}_{e}$, and the DSNB. Within this energy range, the QE and $2p2h$ processes of neutrino-$^{12}$C interactions are recognized as the paramount. We have examined the model variation associated with the neutrino nucleon initial interactions and final-state information in QE and $2p2h$ processes, and shown that the impact of nuclear ground state, FSI, nuclear deexcitation on inclusive and exclusive cross-sections, and the effect of secondary interactions on neutron multiplicity are ascertained. 

This work provides the first detailed prediction for the $\nu_{\rm atm}$ NC backgrounds in the energy range below 100 MeV. The event rate of the IBD-like NC background with a visible energy below 12 MeV is estimated as (2.1 $\pm$ 0.7) kt$^{-1}$ yr$^{-1}$.
The dominant source of uncertainty is attributed to variations in models of the nuclear effect.
The NC background accounts for around 0.2\% for the reactor $\bar{\nu}_{e}$ IBD signal at JUNO~\cite{JUNO:2024jaw}, remaining as one of the non-negligible backgrounds for precise measurements. This prediction has been utilized in determining the neutrino mass ordering with reactor $\bar{\nu}_{e}$. 
Additionally, this prediction proves to be useful in searches for geo-neutrinos. Regarding the substantial interest in the first detection of the DSNB in the energy range between 12 and 30 MeV, we have updated the prediction of the corresponding $\nu_{\rm atm}$ NC background by using advanced data-driven models,
in which the event rate is (2.9 $\pm$ 0.4) kt$^{-1}$ yr$^{-1}$. 



Our methodology for predicting the $\nu_{\rm atm}$ NC interactions provides a wide array of application scenarios. 
Firstly, it is instrumental not only in estimating the double-coincident $\nu_{\rm atm}$ NC background, as studied in this work, but also in calculating single, triple-coincident, or other types of $\nu_{\rm atm}$ NC induced backgrounds. Secondly, it can be easily adapted for predicting $\nu_{\rm atm}$ charged-current interactions. Thirdly, while currently applied to LS detectors, this method is also suitable for a variety of experimental settings, such as Hyper-K~\cite{Hyper-Kamiokande:2018ofw}, Theia~\cite{Theia:2019non}, and DUNE~\cite{DUNE:2015lol}, for forecasting their $\nu_{\rm atm}$ related observations.
Finally, we would like to note that future measurements from these experiments could potentially test and refine the current model predictions, including aspects of primary neutrino-nucleus scattering and secondary interactions in the detectors. We plan to explore this intriguing study in future publications.

\section*{Acknowledgements}
The authors would like to thank Wan-Lei Guo and Xian-Guo Lu for helpful discussions. We also thank Davide Basilico and Liang Zhan for carefully reading the manuscript and useful comments. 
This work is supported in part by the National Key R\&D Program of China under Grant No. ~2024YFE0110500, by National Natural Science Foundation of China under Grand Nos. ~12405125, ~12075255, ~11835013, ~12125506.

\end{document}